\documentclass[sigconf]{acmart}
\usepackage{epsfig}
\usepackage{setspace}
\usepackage{subcaption}
\usepackage{url}
\usepackage{hyperref}
\usepackage{microtype}
\usepackage{graphicx}
\usepackage{tabularx}
\usepackage{color}
\usepackage[labelfont=bf]{caption}
\usepackage{booktabs} 
\usepackage{lipsum}
\usepackage{tabularx,ragged2e,booktabs}
\usepackage{float}
\usepackage{amssymb}
\usepackage{mathtools}
\usepackage{xcolor}

\newcommand{\jafix}[1]{{}}
\newcommand{\jffix}[1]{{}}
\newcommand{\smsfix}[1]{{}}
\newcommand{\fix}[1]{{}}

\newcommand{\ner}{{NER}}
\newcommand{\nerlong}{{named entity recognition}}

\newcommand{\nes}{{NES}}
\newcommand{\neslong}{{named entity search}}
\newcommand{\NESlong}{{Named Entity Search}}

\begin{abstract}
Traditional information retrieval treats named entity recognition as a pre-indexing corpus annotation task, allowing entity tags to be indexed and used during search. Named entity taggers themselves are typically trained on thousands or tens of thousands of examples labeled by humans.

However, there is a long tail of named entities classes, and for these cases, labeled data may be impossible to find or justify financially. We propose exploring named entity recognition as a \emph{search} task, where the named entity class of interest is a query, and entities of that class are the relevant ``documents''. What should that query look like? Can we even perform NER-style labeling with tens of labels? This study presents an exploration of CRF-based NER models with handcrafted features and of how we might transform them into search queries.
\end{abstract}

\begin{document}

\setcopyright{acmlicensed}
\acmConference[LND4IR '18]{ACM SIGIR 2018 Workshop on Learning from Limited or Noisy Data}{July 12, 2018}{Ann Arbor, Michigan, USA} 
\acmYear{2018}
\copyrightyear{2018}
\acmDOI{}
\acmISBN{}
\acmPrice{}


\title{Named Entity Recognition with Extremely Limited Data}
\author{John Foley, Sheikh Muhammad Sarwar, and James Allan}
\affiliation{Center for Intelligent Information Retrieval}
\affiliation{CICS, University of Massachusetts Amherst}
\affiliation{}
\email{[jfoley, smsarwar, allan]@cs.umass.edu}

\maketitle

\section{Introduction}
Consider a reporter, researcher, or policy maker interested in the Syrian refugee crisis, who has found an article discussing how Malta, as a country, has been impacted by the influx of refugees. Suppose this user now wants to find information that details how other island-nations in the Mediterranean have been affected. One technique that could help satisfy this information need is \nerlong{} (\ner{}). If a tagger had marked \emph{countries} and \emph{islands} or (better yet) \emph{Mediterranean islands} and they were indexed by the system, the searcher could construct a query targeting documents that had mentions of refugees and islands (or Mediterranean islands). Why, though, would \emph{islands} have been among the obvious taggers to construct in advance? Would someone bother to build a tagger for \emph{Mediterranean islands}?

Collecting this kind of data may be difficult or impossible for a domain expert. Sequence labeling tasks like this require fine-grained labels and experienced annotators.


Query-log or knowledge-base analysis would allow one to derive a set of entity classes that are of interest, but entity types would have the same long-tail problem as language: there would always be classes missing. Even without approaching the long-tail, it is fairly simple to come up with entity classes that do not quite fit the traditional four-class paradigm (PER, LOC, ORG, other), such as \emph{musical genre}, which is obviously none of the above, or \emph{cigarette brand} which while being related to company or ORG is definitely a distinct entity class. As an experiment, we labeled a number of entities that would have been helpful for TREC QA questions, and found a wide variety of entities, from the common \textit{country} and \textit{city} to the rare: \textit{black panther members}, \textit{airlines}, \textit{sub-atomic particles} and \textit{comets}. Coming up with a comprehensive set of classes a priori seems destined to fail. 

Instead, we propose an approach that treats named entity detection as a \textit{retrieval} task, one performed at query time rather than in advance while indexing. We do not propose this as a replacement for \ner, but as something to be used for an ephemeral or contextual class of entity, when it does not make sense to label hundreds or thousands of instances to learn a classifier. We will show that it is possible to achieve reasonable effectiveness using this approach, though it does require a substantially larger index. We will further show that this approach can respond rapidly, an important validation since we are moving a document pre-processing step to query time.

We are interested in \textit{transforming} the recognition task into a search task and we refer to the task as Named Entity Search (\nes{}). We do that by considering each word occurrence as a document and the words' features as ``terms'', allowing us to use traditional search engines in order to score words in response to a query, where the query is the entity class of interest. 

In this work, we explore four core research questions:
\begin{enumerate}
    \item How do we translate CRF-based NER models with handcrafted features into a retrieval model where tokens are documents? (See \S\ref{sec:model} and Table~\ref{table:asquery}.) By focusing on models with handcrafted features, our users could potentially edit or provide feedback on the features in our short queries~\cite{IR-1124}.
    \item Can we produce a reasonable ranking over tokens for the user to interact with sentences when we only have a handful of labels to create an NER-classifier? (See Figure~\ref{fig:learning_curves}.)
    \item Can we make our NER models small enough that they can be efficiently executed by a search engine without loss of effectiveness? (See Figure~\ref{fig:qasampling}.)
    \item What typical handcrafted NER features are most useful for our ad-hoc variant of NER? (Results in Table~\ref{table:features}.)
\end{enumerate}
We present experiments and analysis for these research questions, and begin by discussing related work.

\section{Related Work}\label{related-work-section}

Named entities have a long history of being important and useful for information retrieval tasks. Recently, work has focused on connecting entities present in text to knowledge bases to allow for improved reasoning~\cite{dalton2014entity,xiong2015esdrank}. Entities have also been at the center of a large amount of research, including the entity recognition and disambiguation challenge~\cite{carmel2014erd}, the TREC temporal summarization track~\cite{aslam2013trec}, the TREC knowledge-base population track~\cite{mcnamee2009overview}, and the entity ranking tracks at INEX~\cite{demartini2010overview} and TREC~\cite{balog2010overview}. Entities, specifically people with emails, were also the highlight of the expert-finding task within the TREC enterprise track~\cite{craswell2005overview}. 
Almost all of these problems and domains depend on entity recognition as a document-enrichment step.

In natural language processing, \nerlong{} (\ner) and part-of-speech tagging (POS) are considered examples of the sequence labeling problem. One of the common evaluations for this task is the CoNLL-2003 shared task~\cite{tjong2003introduction}, which introduced English and German datasets derived from expert-labeled news articles. Successful \ner{} systems use a large number of features, including gazetteers, case information, word cluster ids, and part-of-speech tags. A 2007 survey by Nadeau and Sekine~\cite{nadeau2007survey} provides more information about the origins of this task and traditional approaches. A recent work that discusses the challenges of generalization for modern NER approaches~\cite{augenstein2017generalisation} may be an additional resource in lieu of a more recent survey.

\subsection{Recent Approaches in NER}

Most state-of-the art NER approaches are now based on neural networks and deep learning. While using neural networks for \ner{} is not new (one of the CoNLL 2003 submissions used a LSTM network~\cite{tjong2003introduction,hammerton2003named}), the availability of more memory and more data has made these approaches the dominant research direction. Recent advances in deep neural networks have led to a reduction in the number of hand-tuned features required for achieving state-of-the-art performance. 

However, since we expect our task to have very few labels, and we hope to minimize the complexity of our query models. Our work includes the ``distributional similarity features'' included in Stanford NLP, which are Brown clusters~\cite{brown1992class,liang2005semi}. In future work, we hope to explore explicitly learning neural representations from unlabeled data, e.g. Lample et al~\cite{lample2016neural}.



\subsection{Many-classed NER}

Some work explores more fine-grained representations of entities, namely a work by Lee et al.\ which explores a model for identifying named entities as a single step and classification into 147 different categories as a secondary step~\cite{lee2006fine}. More recently, Ling and Weld present \textsc{FIGER}, a system for fine-grained entity recognition of 112 tags, trained using data collected from Wikipedia links used a similar approach: a CRF for segmenting into B (beginning), I (intermediate), and O (outside) tags, and a multiclass classification task on top~\cite{ling2012fine}. Lin et al.\ assign more than 1000 types to noun-phrases to help classify out-of-knowledge-base entities~\cite{lin2012no}. In order to be as general as possible, we do not consider filtering to noun-phrases or classified spans in this work, but that is a promising approach to increasing efficiency in future work.

\subsection{List Completion and QA}
Two existing tasks are similar to \nes{} in terms of their goal: generating a list of entities based on user intent. However, neither explores a limited-data setting.

\subsubsection{Question Answering Systems} First, question answering systems often leverage \nerlong{} as a method of collecting potential answers~\cite{molla2006named,lee2006fine}, for both factoid QA as well as list QA~\cite{trecqa05,trecqa06}. However, historically, some of the best systems required intensive data resources that do not make sense for our ad-hoc task. The best system from 2007 used coverage from the lists available in Wikipedia, ahead-of-time construction of target NER classes, and sophisticated textual analysis based on logic systems~\cite{moldovan2007lymba}. Another consistently high-ranked system used a tiered response system that first checked a database built from ``info boxes,'' then Wikipedia lists and tables, then a set of data from lexicons, and even web search hit counts~\cite{hickl2007question}. Often, systems targeting factoid queries merely repeatedly executed their factoid method. Because of our reluctance to limit potential answers to those in external resources and our desire to incorporate interactivity we do not compare directly to any TREC-QA systems in this work. However, we do make use of the list questions in the TREC 2005 and 2006 Question Answering track~\cite{trecqa05,trecqa06}.

\subsubsection{List Completion} List completion exists in various forms, with some work inspired by the no-longer-running Google Sets~\cite{dalvi2011entity}. However, this work presumes that data mining methods will be effective: that entities of interest occur in HTML or Wikipedia lists with some regularity. A similar task was explored at the INEX list completion task \cite{demartini2010overview}, which was based on selecting the best answers from a fixed Wikipedia corpus. In the language of existing List Completion approaches, we could call our task \textit{list completion over unstructured text}, but we are not aware of any similar work.


\subsection{Active Learning}

Active Learning is a broad field that arose from the desire to have machine learning algorithms select a subset of the training data to learn the best classification function, but the tools and techniques in this field are applicable to interactive labeling tasks as well. Settles presents a comprehensive survey of this subfield~\cite{settles2010active}.

In the realm of \ner, active learning has specifically been used to minimize human annotation efforts. Shen et al.{} show that only 20\% of labels are truly required to achieve strong performance with a good selection function~\cite{shen2004multi}. Rather than minimizing human interactions to achieve a final high-accuracy classifier, our \neslong{} task focuses on \emph{extremely} limited interaction and high-precision effectiveness.

\section{Methods}\label{sec:model}

The core of our interactive \NESlong{} is the ability to leverage user-feedback in order to quickly and interactively construct useful named entity taggers. We briefly discuss conditional random field, as the dominant approach to non-neural NER as it is more suitable for our fewer label, interactive setting. And then we talk about the user interaction model.

\subsection{CRF as a Retrieval Model}

The retrieval model specifies how retrievable items (tokens represented by features) are ranked in response to the query (weights of features). Based on the success of CRFs, we start with its calculation of the probability that a token $t$ has a label $y_t$. We use the notation of Sutton and McCallum~\cite{sutton2010introduction}: 

\[
p(y_t | x) = \frac{1}{Z} \exp \left( \sum_{k=1}^{K} {w_k f_k(y_t, y_{t-1}, x_t)}\right)
\]

To make this useful for search, we first apply some standard IR transformations to this model. We intend to rank tokens consistent with this probability rather than calculating it directly. 

Because we are interested in ranking, rather than exact probabilities, we can perform a number of typical rank-safe transformations: removing the $\frac{1}{Z}$ normalization, and taking the log of each side. We cannot remove the dependency of $y_t$ on $y_{t-1}$ with pure mathematics, so we turn to experimentation to determine whether we can simplify this expression further and to quantify the loss of accuracy.

We also validate empirically that with linear CRF models, especially with fewer labels, the transition probabilities are of little benefit. Table~\ref{table:asquery} shows the results of downgrading a full CRF to doing a single class at a time and then removing transition probabilities on a standard NER dataset. By ignoring the transition ans sequence passes of learning and training classes independently, we have achieved a linear query model and measured the difference between that and the more typical CRF.

\[ p(Q | x) = \sum_{k=1}^{K} {Q_k f_k(x)} = \vec{Q} \cdot \vec{f}(x) \]

This means that we treat each token in a collection as a document, and it is represented by a bag of traditional NER features (as extracted by Stanford NLP~\cite{manning2014stanford}). This means that we can effectively execute our NER models with an IR system that supports this sort of vector-space model.

\begin{table}[ht]
  \setlength\tabcolsep{3pt}
  \caption{Classification Results on our NER Dataset. {\rm Measures are the $F_1$ of token classification experiments to explore the effect of our new assumptions.}}\label{table:asquery}
  \centering
  \begin{tabular}{l|rrrr|rr}
    System       & PER  & LOC  & ORG  & MISC & Micro & Macro \\ \hline
    Full CRF     & 86.0 & 85.6 & 91.1 & 61.9 & 86.5 & 81.1 \\
    Query independence & 85.2 & 87.3 & 90.5 & 59.2 & 86.3 & 80.6 \\
    Token independence & 84.5 & 86.4 & 89.8 & 63.0 & 85.9 & 80.9 \\
  \end{tabular}
  \vspace{-1em}
\end{table}

\subsection{User Interaction}

In early feedback, we discovered that returning tokens or spans to the user was not sufficient to understand whether the entities being discovered were correct, so we built our system around a sentence-feedback model, where users are presented a sentence at a time (which contains the highest ranked tokens) and users label the entire sentence before the system re-ranks and presents them with the next best sentence. In this way, a user works to solve their problem (collecting a list of entities) while generating a small amount of training data suitable for NER software~\cite{okazaki2007crfsuite}.


\section{Datasets}
\label{sec:datasets}

We explore questions related to \nes{} using a well-labeled \ner{} dataset on which a classification task makes sense. Our fundamental research questions regarding whether \nes{} is practical and effective are done using a novel QA-List dataset.

\subsection{\ner} We leverage the English dataset from the CoNLL-2003 shared task. Table~\ref{table:stats} provides some statistics about that collection. It consists of 946 documents for training, 231 documents for validation, and 231 documents for testing. Our approach does not need to train parameters, so we did not use the validation set. In order to compare to previously published results, we choose to ignore that portion of the data rather than mix it into the training or test sets. 

\subsection{QA-List} 
In order to explore research questions that surround sparse entity types, we have adapted a dataset from the TREC Question Answering Track. We select data from the 2005 and 2006 challenges because they operate on the same dataset (AQUAINT) and the full set of annotator judgments are available for all surface forms reported by all participants in the challenge. Some statistics about AQUAINT in comparison to the smaller NER dataset are presented in Table~\ref{table:stats}.

\begin{table}[ht]
  \caption{Comparison and Summary of Datasets.$\ast$ \\
{\rm  We present a comparison of sizes here to emphasize that the na{\"i}ve storage cost of our new task (storing per-token features as documents) expands the size of the collection, motivating our look into efficiency concerns.}}\label{table:stats}
  \begin{minipage}{\columnwidth}
  \centering
  \begin{tabular}{lrr}
    Collection       & NER & QA-List \\ \hline
    Source           & CoNLL'03 & AQUAINT \\
    \#docs           & 1408     & 1,088,791 \\
    \#words          & 203,621  & 488,789,688 \\
    \#uniq words     & 20,386   &  1,113,614 \\
    Inv \& Fwd Index & 1.8M      & 2.8G \\
    NER Features     & 276M     & 495G \\
    NER Features Index   & 95M      & 158G \\
    CRFSuite Tagging Time & 2 seconds & 2 hours \\ \hline
  \end{tabular}
  \end{minipage}
\centering {\footnotesize $^\ast$ G=$2^{30}$ bytes, M=$2^{20}$ bytes, as calculated by \texttt{du -csh}}
\end{table}

Since we are explicitly curating this data to use less-common entity types, we preprocess it to skip the 25 queries that are explicitly seeking lists of countries and the 4 queries that are seeking lists of cities. We also drop queries that have 3 or fewer answers because we want to show learning curves for this task. After preprocessing, we have 66 queries from 2005 and 60 queries from 2006. 
Some examples can be found in Table~\ref{table:examples}.  

We limit our processing to the provided ranking created by the track organizers. While this limits our recall slightly (there are positive judgments outside these rankings), this prevents our technique from pulling up large numbers of unjudged tokens, and makes our experiments more computationally efficient. 
In the future, we hope to fully adapt this dataset and collect the additional judgments.

\begin{table}[h]
\caption{Examples of QA-List Questions}\label{table:examples}
    \begin{tabularx}{\columnwidth}{XrX}
        Title & QID & Items \\ \hline
        Ben \& Jerry's unusual flavors & 172.7 & Chubby Hubby, Phish Food, Cherry Garcia, $\ldots$ \\
        Mammals cloned & 197.4 & rabbit, cow, mouse, pig, $\ldots$ \\
        Vaccines for Avian Flu & 166.6 & Relenza, Tamiflu, amantadine, RWJ-270201,  $\ldots$\\
        Boxers who defeated \mbox{Foreman} & 77.7 & Muhammad Ali, Shannon Briggs, Evander Holyfield, Tommy Morrison,  $\ldots$\\
        Names of Meteorites & 84.7 & Lucky 13, ZAG, ALH84001, Leonid, SNC, $\ldots$
    \end{tabularx}
\end{table}

\subsection{Unique Average Precision (uAP)}

For our person query (PER), we do not want to claim high effectiveness if the only person it tags in the top ranks is ``Barack Obama'', even if our system is technically correct about the class of all those repeated instances. We define \textit{unique average precision (uAP)} to be standard average precision, but only after \textit{removing} all subsequent mentions of any entity.  Unlike evaluations of retrieving novel material~\cite{waterloo08novelty}, uAP \textit{ignores} subsequent occurrences of both relevant and non-relevant mentions rather than treating them all as non-relevant.
Although $F_1$ is typically used for NER, it is unsuitable for \nes{} because it is a set-based measure and we care about the actual ranking. (We do use $F_1$ when comparing \ner{} results.)

This notion of unique-AP is our way of encoding the ``distinctness'' that was done by human assessors in the TREC-QA list track: they marked arbitrary instances as the distinct instance, and counted it as a recall point for AP if and only if it was both correct and distinct. Our measure does the same, except at the token level, and automatically.

\section{Interactive \nes{} for challenging entity classes}\label{results:qalist}

In this section, we use our derived QA-List dataset (Section~\ref{sec:datasets}) in order to explore learning curves on a more realistic task.

While the active learning literature~\cite{settles2010active} has thoroughly explored methods for classifiers to select instances for labeling, we have a slightly different twist on this task. Instead of collecting labels through active learning while running the full classifier, we want to see if our approximations still allow for a strong learning curve.

\subsection{Is Interaction Helpful?}

We explore three interaction models for the \nes{} problem of labeling the entities of interest. The first model is our own, an interactive model that labels the top-ranking sentence, and immediately feeds it back to create a new query and thus a new ranking. The second model is one that labels sentences in the top-ranked documents (for the NER dataset, we use collection order in lieu of rank). The final model is that of a more traditional NER labeling: an assessor chooses random sentences. In the QA-List dataset, our random baseline is built on document pools, since in expectation a truly random selection of sentences from AQUAINT would lead to no positive labels over hundreds or thousands of iterations. Our final baseline "Unsure" is derived from active learning: it chooses an instance based on how unsure the classifier is about it, with the intuition that these instances will be most informative to the learning algorithm.

\begin{figure*}

  \begin{subfigure}{0.45\textwidth}
  \centering
  \includegraphics[width=2in]{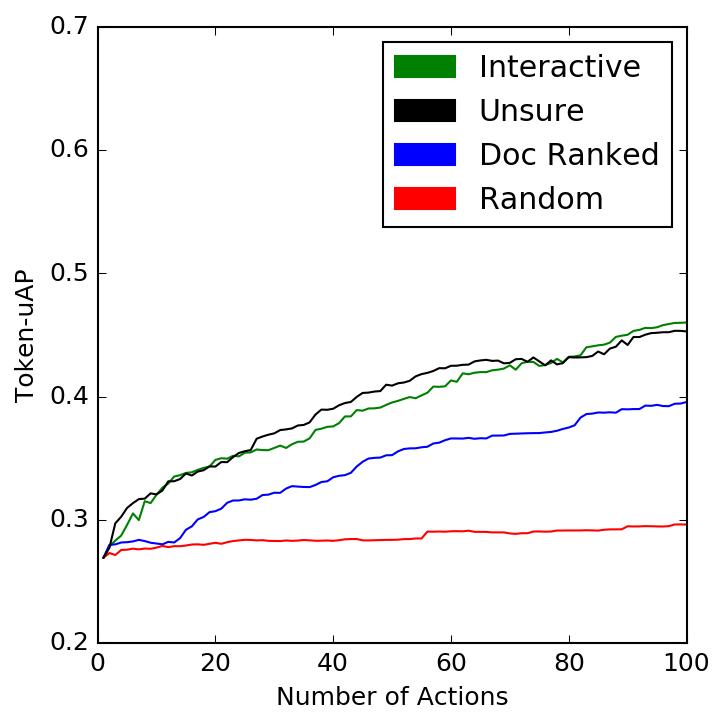}
  \caption{QA-List dataset.}\label{fig:interactive_methods}
  \end{subfigure}
  \begin{subfigure}{0.45\textwidth}
  \centering
  \includegraphics[width=2in]{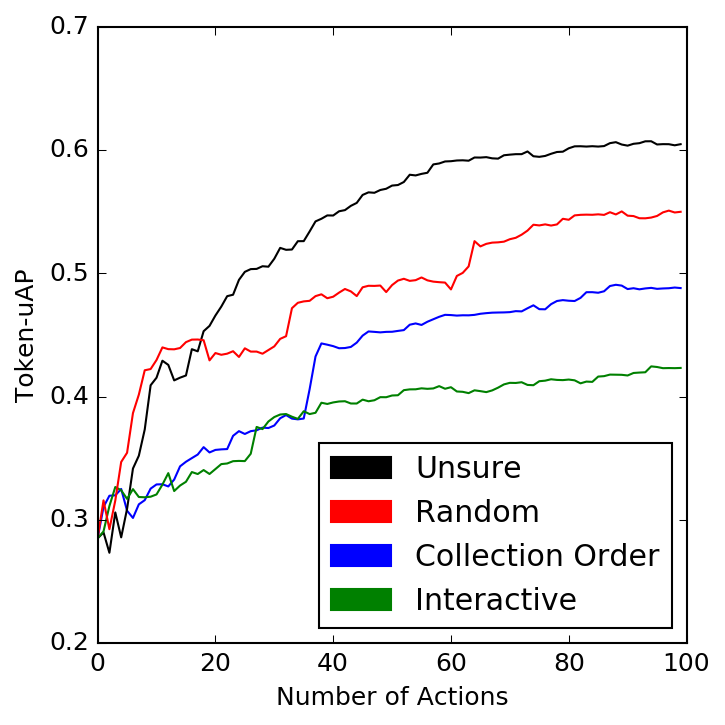}

  \caption{NER dataset. }\label{fig:interactive_methods_conll}
\end{subfigure}
\caption{{\bf Interactive effectiveness on our two datasets.}\\
  {\rm The NER dataset (right) is deeply labeled, but is less informative for our desired problem as a result of the entity classes being too frequent and too simple. 
  On our new QA-List dataset, however, we can see that the unsure and interactive baselines are equally competitive, which validates our new, interactive approach to NER for more difficult datasets and sparse labels.}}
  \label{fig:learning_curves}
  
\end{figure*}

Interactivity curves for our QA-List dataset are presented in Figure~\ref{fig:interactive_methods}. It shows the change in effectiveness (uAP) as the query changes via interaction. Selecting random documents is useless and is easily out-performed by a strategy that labels sentences from high-ranking relevant documents. 

Similar curves over the NER dataset are presented in Figure~\ref{fig:interactive_methods_conll}, but shows nearly the opposite effect. This corpus is unusual in that it has an extremely high concentration of named entities, so that a random sentence is very likely to contain one. As a result, randomly selected text creates a recall-enhancing effect. 

We believe that the QA-List dataset captures our target problem more clearly: we expect users to have information needs regarding sparse, ``tail'' entities. However, these results point up the need for retrieval models that can find diverse entities to boost recall.

\subsection{Is this efficient and effective?}

In order to demonstrate our model's possibilities, we must be proposing something feasible. 
This is important because we can evaluate our smaller models with much less CPU-time over a much larger dataset than traditional CRF systems (see Table~\ref{table:efficiency}), which is important for any sort of interactivity on real, modern datasets. From an opportunity cost perspective, the inverted list algorithm beats direct scoring for a corpus of this size as long as it executes in less than an hour, but those times are still unreasonable for users.

\begin{table}
    \caption{{Query Efficiency vs. Model Size} {\rm This table shows the trade-off between the interactive IR-approach to model evaluation and the offline-NLP approach over the full AQUAINT corpus.}}\label{table:efficiency}
    \centering
    \begin{tabular}{lrr@{.}lr}
        Algorithm & $|Q|$ & \multicolumn{2}{c}{Median Time(s)} & Best Unit \\ \hline
        Lucene & 1 & 0&135 & $<1$ sec \\
        & 5 & 0&319 & $<1$ sec \\
        & 10 & 0&499 & $<1$ sec \\
        & 50 & 5&45 & $<10$ sec \\
        & 100 & 22&4 & $<1$ min \\
        & 250 & 219&0 & 3-4 min \\
        & 500 & 566&0 & 9-10 min \\
        & 1000 & 1700&0 & 30 min \\
        & 5000 & 12200&0 & 3-4 hours \\ \hline
        CRFSuite & Full & 7200&0 & 2 hours \\
    \end{tabular}
    \vspace{-1em}

\end{table}

\begin{figure}[]
    \centering
    \includegraphics[width=2in]{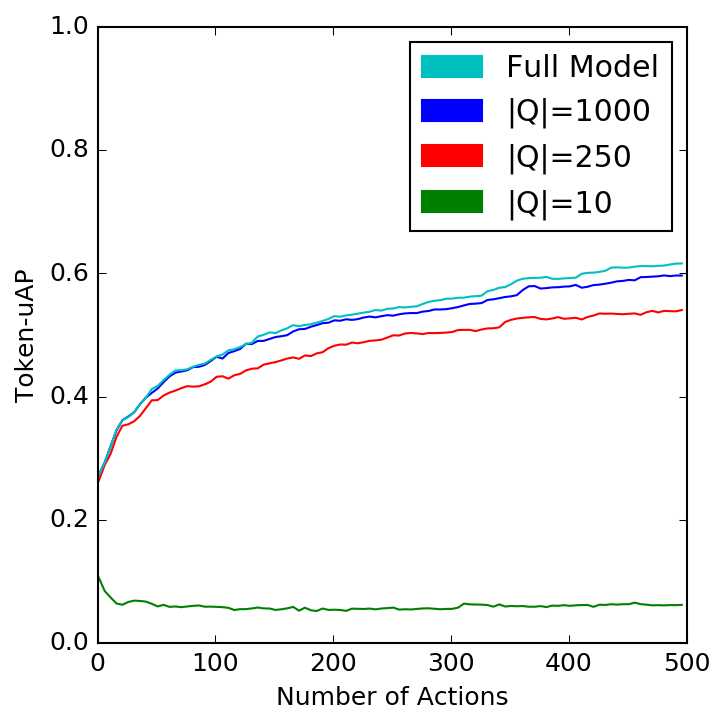}
    \caption{Effectiveness on QA-List with limited features for efficiency. {\rm On this dataset, the learning curve for models limited to 1000 features is nearly indistinguishable from the curve with full models. Limiting model size does not result in a significant performance drop within a few hundred interactions.}}\label{fig:qasampling}
\vspace{-1em}
\end{figure}

In our challenging QA-List task, we have been able to show that the learning curves of down-sampled models are insignificantly different than the full models. When we consider this in conjunction with our efficiency results, this means that we could execute our interactive system on the full AQUAINT dataset in just a few CPU-minutes rather than hours. See Figure~\ref{fig:qasampling} for representative curves. If we limit ourselves to a few hundred of the most important features, we get nearly all of the benefits without using hours of CPU-time.

\subsection{What features are most effective?}

We studied the top-10 features selected by each model (weighted by their occurrence across models and by those models' uAP score), and found the top 5 most important features by domain. These results are presented in Table~\ref{table:features}.

\begin{table}
    \centering
    \caption{The top features in our analysis.}
    \vspace{-1em}
    \label{table:features}
    \begin{tabular}{cll}
        Rank & NER & QA-List \\ \hline
        1 & Brown Cluster Curr & Words to the Right \\
        2 & POS Tag            & Words to the Left \\
        3 & Char N-grams       & Char N-grams \\
        4 & Brown Cluster Next & Word Shape Next \\
        5 & Words to the Right & Word Shape Prev,Curr 
    \end{tabular}
\end{table}

The most useful features for the NER domain were word clusters~\cite{brown1992class,liang2005semi} 
trained on the RCV1 corpus provided pre-trained by Stanford NLP~\cite{manning2014stanford}. 
We briefly explored training\footnote{\url{https://github.com/percyliang/brown-cluster}} our own word clusters on the AQUAINT corpus but we observed no significant difference in learning curves. We hope to explore more distributional similarity features in the future.

\section{Conclusion}

In this study, we explore the possibility of viewing a traditionally offline pre-processing and token classification task as an exploratory search task to deal with extremely limited data. In so doing, we introduce a novel problem that could help analysts, reporters, and other expert users understand and solve their information needs.

We explore effectiveness, modeling, and efficiency while focusing on handcrafted features rather than neural models so that we can explore human-edited queries in the future. We conclude that our approach will be feasible for expert use. In the future, we hope to fuse more research from the active learning, NLP, and IR fields to develop new approaches to this task.

\section*{Acknowledgements}
This work was supported in part by the Center for Intelligent Information Retrieval and in part by NSF grant \#IIS-1617408. Any opinions, findings and conclusions or recommendations expressed in this material are those of the authors and do not necessarily reflect those of the sponsors.

The authors would additionally like to thank Jiepu Jiang, who helped us to extract the NLP features at the core of these experiments.


\bibliographystyle{abbrv}
\bibliography{ms}
\end{document}